\documentclass[ amsmath, amssymb, aps, 12pt,superscriptaddress]{revtex4-1}

\usepackage{graphicx}
\usepackage{dcolumn}
\usepackage{bm}
\usepackage{hyperref}
\usepackage{xcolor, soul}
\usepackage{braket}

\begin{document}

\title{Spin coherence mapping of NV centers in diamond}

\author{Lucas Nunes Sales de Andrade}
\affiliation{ Instituto de Física de São Carlos, Universidade de São Paulo (IFSC-USP),
\\ Caixa Postal 369, CEP 13560-970, São Carlos, SP, Brazil. \\ ( srmuniz@ifsc.usp.br -- https://orcid.org/0000-0002-8753-4659 ) }

\author{Charlie Oncebay Segura}
\affiliation{Facultad de Ciencias, Universidad Nacional de Ingeniería, Lima, Peru.}

\author{Sérgio Ricardo Muniz}
\affiliation{ Instituto de Física de São Carlos, Universidade de São Paulo (IFSC-USP),
\\ Caixa Postal 369, CEP 13560-970, São Carlos, SP, Brazil. \\ ( srmuniz@ifsc.usp.br -- https://orcid.org/0000-0002-8753-4659 ) }

\begin{abstract}
\vspace{7mm}
In recent years, nitrogen-vacancy (NV) color centers in diamond have become excellent solid-state quantum sensors due to their electronic spin properties. Especially for their easy optical initialization and detection, together with their very large spin coherence at room temperature. Many studies have reported their use for sensing temperature, strain, electric fields, and mainly magnetic fields. Here, we show how to build a two-dimensional map of the electronic spin coherence of an ensemble of NV centers in ultra-pure diamond, using an optical imaging protocol combined with microwave pulses relying simply on a regular CCD camera. 
\end{abstract}
\maketitle

\section{Introduction}
One of the most exciting features of the NV center is its long spin coherence time at room temperature. Combined with the easy optical reading of single spin states, it makes the NV center in diamond a promising platform for quantum technologies. Its primary use has been as a magnetic field sensor \cite{acosta2009diamonds, nanoscale, biomag, ultrasensitive, Tetienne2017}, but it has also been used as a sensor for electric fields \cite{electric}, temperature \cite{nanothermo, thermometer2020}, and also for quantum information processing \cite{quantum}.

Here, we present a new protocol using a thin layer of NV centers and a CCD camera to perform a pixel-by-pixel analysis that allows one to do a two-dimensional (2D) mapping of the spin coherence of our device. It uses a simple microwave pulse sequence and an optical imaging protocol adapted from our previous work on wide-field vector magnetometry \cite{charlie2021, lucas2021}.

\section{Methods}

\subsection{Sample preparation}
An ultra-pure (CVD) diamond plate was thinned and polished to a size of $2\times 2\times 0.1\,\mbox{mm}^3$. To create NV centers near the top surface, the sample was irradiated with $^{15}\mbox{N}$ ions at $5 \,\mbox{keV}$ and fluence of $1\times 10^{13}\,\mbox{cm}^{-2}$. Under these conditions, a nearly homogeneous layer of nitrogen is created at around $8$ to $16$ nm below the surface \cite{Tetienne2017,Haque-Sumaiya}, in addition to a number of vacancies produced by collision along the path of the nitrogen ions. After the ion implantation, the sample was annealed at $800 ^{0}\mbox{C}$ for a couple of hours, in vacuum, to allow the migration and trapping of the vacancies to the implanted nitrogen.

\subsection{Optical pumping and state initialization} \label{Section:A}
\begin{figure}[thb]
\centering
\includegraphics[scale=0.75]{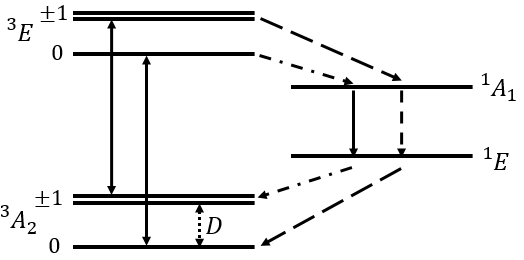}
\caption{Electronic structure of the NV center with relevant transitions. The ground triplet state $^3A_2$ is driven to the excited triplet state  $^3E$ via radiative path (represented by solid arrows). The decay is spin-dependent, where dashed arrows represent strong non-radiative decays, and point-dashed arrows indicate weak non-radiative decays. The dotted arrow indicates the microwave transition.}
\label{nvstruct}
\end{figure}

The relevant electronic levels of the NV center are the triplets $^3A_2$ ground states and $^3E$ excited states, plus two metastable states $^1A_1$ and $^1E$, as depicted in Fig. \ref{nvstruct}.

Initially, at thermal equilibrium at room temperature, the populations of the triplet ($\ket{0}$ and $\ket{\pm 1}$) ground states are approximately equal, following the Boltzmann factor. In this situation, there is no spin polarization.

When the system is excited by a green laser ($532\,\textrm{nm}$), the ground state is driven to the excited state through an allowed optical transition with a zero-phonon line (ZPL) at $637\,\textrm{nm}$ \cite{thesepreez}. The decay times are strongly spin-dependent and involve non-radiative processes. For instance, the excited $\ket{0}$ state most often decays directly to the ground state, while the states $\ket{\pm 1}$ often cross over into the $^1A_1$ and $^1E$ states \cite{doherty2013nitrogen}, see Fig. \ref{nvstruct}. From there, it eventually decays back to the ground state.  However, since the overall decay times are quite different, after several excitation cycles, the system accumulates in the $m_s=0$ ground state \cite{doherty2013nitrogen}, effectively leading to a process of optical pumping that initialize the system into the $\ket{0}$ ground state.

The relative change in intensity of the photoluminescence (PL) signal due to the direct relaxation ($^3A_2 \leftrightarrow\,^3E$) is used for determining the spin state of the NV center, because the process of crossing through the intermediate metastable states, $^1A_1$ and $^1E$, requires more time than the direct relaxation \cite{Acosta_PRB}, leading to a change in the rate of photons per unit time.

\subsection{Microwave interaction}
In the presence of a magnetic field, due to the Zeeman interaction, the $\ket{\pm 1}$ states, previously degenerated, will have an energy difference given by $\Delta E = 2 \gamma_e B_z$, where $\gamma_e \approx 28\,\textrm{MHz/mT}$ is the electron's spin gyromagnetic ratio, and $B_z$ is the magnetic field component on the NV axis. Therefore, the effective ground state Hamiltonian, in the $S_z$ basis, can be simplified to \cite{thesecharlie}

\begin{equation}
    H_{gs} = DS_z^2 + \gamma_e B_z S_z, 
    \label{hamilt}
\end{equation}
where $D \approx 2.87 \,\textrm{GHz}$ is the Zero Field Splitting (ZFS) constant. Here we are neglecting terms of order $1 \,\ \textrm{MHz}$ or lower, to keep the description simple.

Applying a weak oscillating magnetic field (microwave (MW)) perpendicular to the main field $B_z$, we are able to make transitions: $\ket{0} \rightarrow \ket{1}$ (if $\omega_{mw} = D + \gamma_e B_z $) or $\ket{0} \rightarrow \ket{-1}$ (if $\omega_{mw} = D - \gamma_e B_z $), where $\omega_{mw}$ is the MW frequency. These transitions cause a reduction in the photon count rate, and as discussed in section \ref{Section:A} the photoluminescence signal is related to the spin state of the system. This technique is known as Optically detected magnetic resonance (ODMR).

In a simplified description, one can write the projection of a general state $\ket{\psi}$ onto the state $\ket{0}$ in terms of a PL signal $\mathcal{S}$ given by \cite{lucas2021,thesecharlie}

\begin{equation}
    \mathcal{S}(t) \propto e^{-\left(t/\text{T}_2^*\right)} \sin^2 \left(\frac{\Omega'_R t}{2}\right),
    \label{rabios}
\end{equation}
where $\Omega'_R = \sqrt{\Omega_R^2 + \Delta^2}$ is the generalized Rabi frequency, and $\Delta=(\omega_{mw}-\omega_0)$ is the detuning between the angular frequency $\omega_0 = D \pm \gamma_e B_z$ and the MW frequency $\omega_{mw}$. 
The parameter $\text{T}_2^*$ represents the characteristic decay time of the spin coherence in the presence of local magnetic inhomogeneities. This is the parameter that we are mainly interested in this paper.

\subsection{Spin Coherence}\label{spincohe}
The time for the spin to return to equilibrium after a perturbation is called the relaxation time. There are two important types of relaxation times: longitudinal relaxation time ($T_1$) and transverse relaxation time ($T_2$).

The longitudinal relaxation time is related to the exchange with the environment (lattice), and for this reason, $T_1$ is also known as spin-lattice relaxation time. It measures the time it takes to return to the equilibrium state from the initially spin-polarized state. On the other hand, the spin-spin interaction leads to the transverse relaxation time or $T_2$. The spin-spin interaction can be either due to electronic or nuclear spins, as in the case of the electronic spin interacting with the nuclear spin-bath of carbon atoms ($^{13} C$). The transverse relaxation time also depends on the inhomogeneities of the local magnetic field, in which case is referred to as $T_2^*$, as well as temperature fluctuations and strain gradients \cite{ultralong}.

\section{Experimental setup}

\subsection{Microwave generation}
The microwave is produced by a frequency synthesizer (Stanford SG384), connected to a fast switch (CMCCS0947A-C2) that receives a TTL signal from a fast digital card (SpinCore, PBESR-PRO-300), with a time resolution of $3.3 \,\ \textrm{ns}$, controlled by a computer. The signal generated by the synthesizer is amplified by a $45\,\textrm{dB}$ power amplifier (Mini-Circuits ZHL-16W-43-S+) and goes to the MW antenna, which has a particular "$\Omega$-shape" tiny-loop to increase the Rabi frequency at the sample. In addition, a circulator (CS 3000) with a $50 \,\Omega$ heavy-duty resistive load is used to protect the system \cite{lucas2021}.

\subsection{Optical Setup}   
The experimental setup can operate in two modes: continuous and pulsed. We use an acousto-optic modulator (AOM) to control and switch between both modes. The AOM alignment was optimized to maximize the green laser's first-order diffraction, at $532\,\textrm{nm}$ (Thorlabs, DJ532-40), and use it as the excitation beam.

The size and position of the beam on the AOM was adjusted to achieve minimum delay ($\approx130\,\textrm{ns}$) and the fastest response time ($\approx35\,\textrm{ns}$), resulting in a total switching time of $165\,\textrm{ns}$. Since the time delay is fixed and stable, we simply adjust the pulse delays to account for it in the timing sequence, obtaining reliable results. 
After the AOM and the iris, a pair of lenses of $25 \,\textrm{mm}$ and $5\,\textrm{mm}$ focal lengths is used to expand and collimate the beam. A dichroic mirror (SemRock, Di02-R561-25x36) reflects the green excitation beam and separates the red photoluminescence emitted by the NV centers from the scattered green light. 

\begin{figure}[t]
\centering
\includegraphics[scale = 0.5]{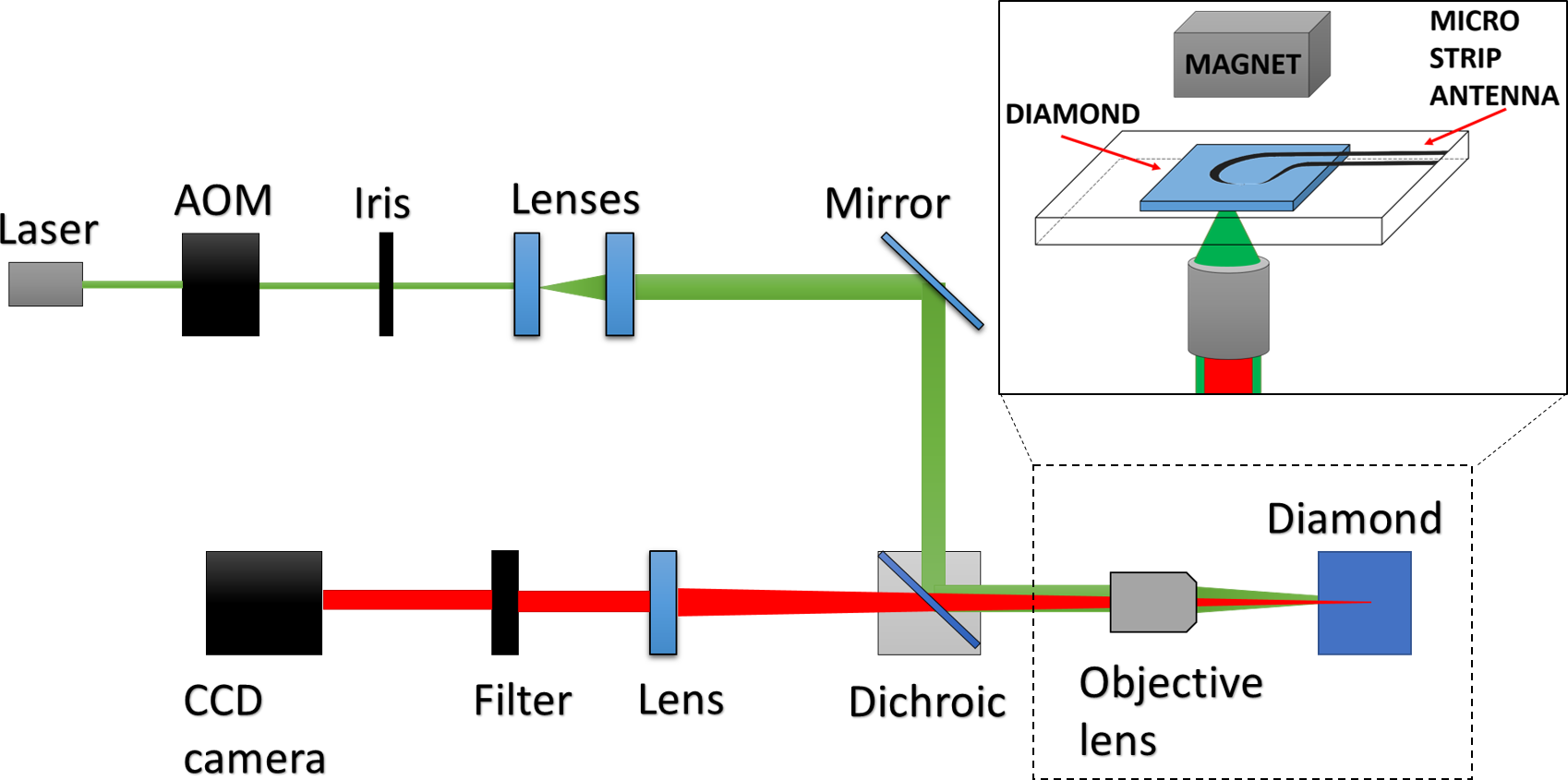}
\caption{The $532\,\textrm{nm}$ laser beam goes through an AOM, and an iris selects the first diffracted order to go through a telescope to enlarge the beam. Next, a dichroic mirror reflects the green light towards an objective lens that focuses it onto the sample and collects the red PL that goes through the dichroic and a spectral filter before being captured by a CCD camera. The boxed region on the top-left illustrates the details (not scaled) of the region near the sample (dashed-box).}
\label{esquema}
\end{figure}

The dichroic reflects light with wavelengths shorter than $561 \,\textrm{nm}$, that goes through the objective lens (Apochromat 50x, NA=0.95), which focuses the light onto the diamond. The same objective collects and collimates the NV fluorescence (PL) that goes through the dichroic and a final lens, forming an image onto a CCD camera (Point Grey, FL3-FW-0S1M-C). An optical filter (ELH0550) was positioned before the camera to block any further green light. Figure \ref{esquema} shows a schematic of the setup \cite{lucas2021}.

\subsection{Optically Detected Magnetic Resonance}
To perform the Rabi oscillations, which give us information about the coherence time $\textrm{T}_2^*$, we need to find the resonance frequencies. Due to the spin-dependent dynamics described in section \ref{Section:A}, when we apply MW at a resonance frequency, it causes a decrease in the PL signal that we detect using a CCD camera. 

\begin{figure}[hb]
\centering
\includegraphics[scale = 0.6]{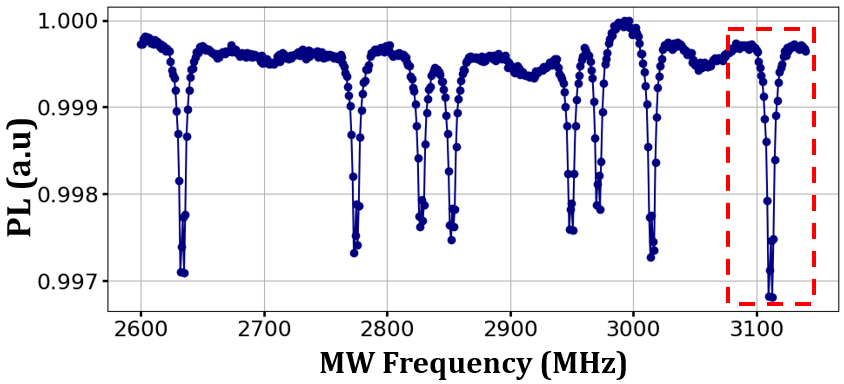}
\caption{Optically detected magnetic resonance (ODMR) spectrum, observed varying the microwave frequency in steps of 1 MHz. The four orientations of the NV axis lead to eight resonances. The chosen frequency of $3111.5\,\textrm{MHz}$ is highlighted in red.}
\label{odmrspectrum}
\end{figure}

In our setup, a permanent neodymium magnet near the diamond provides a bias field $B_z=85.25 \pm 0.07 \,\textrm{G}$, which lifts the degeneracy of the states $\ket{\pm 1}$. Varying the MW frequency, we detect 8 different resonances due to 4 different orientations of the NV axis, as shown in the ODMR spectrum in Fig. \ref{odmrspectrum}. For the experiments described here, we used a resonance peak at $3111.5\,\textrm{MHz}$, which drives the transition $\ket{0}\rightarrow \ket{1}$. 

\subsection{Pulse sequence and data collection protocol}\label{Pulse}
Once the resonant (MW) frequency is fixed, one can use it to control the spin dynamics. First, we apply an optical pulse for $300\,\ \mu\textrm{s}$ to provide the optical pumping to produce the ensemble pure state $\ket{0}$, initializing the system. Then, the laser is turned off, and the spin state is coherently manipulated by applying MW square pulses. 

\begin{figure}[thb]
\centering
\includegraphics[scale=0.5]{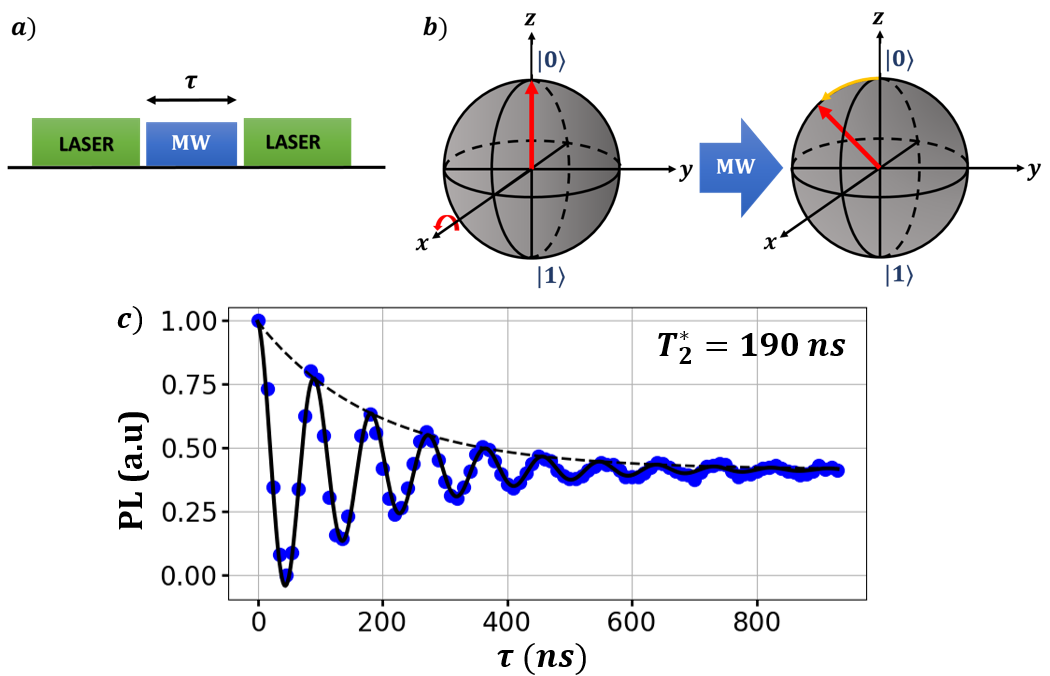}
\caption{a) Pulse sequence to measure Rabi oscillations. b) Representation on Bloch sphere. c) Experimental data (blue) and the fitted line (black) obtained using Eq. (\ref{rabios}), here resulting in $\textrm{T}_2^* = 190\,\ \textrm{ns}$.}
\label{graftrabi1}
\end{figure}

To measure the Rabi oscillations, the MW is applied for various interaction times $\tau$. The PL signal is retrieved by quickly turning on the laser after shutting off the microwave. The laser is kept on for a time long enough to provide the optical pumping that restarts the system to another MW pulse sequence. Fig. \ref{graftrabi1}(a) shows the schematic pulse sequence.

For each time value $\tau$, we collect three different types of images on the CCD camera. For each one, the CCD shutter remains open for $\approx 40\,\textrm{ms}$ to capture as much light as possible while avoiding saturation. The first image (M) has the PL with the complete microwave sequence (shown in Fig. \ref{graftrabi1}(a)) repeated 150 times. The second image (L) has the PL with the same timing sequence and number of repetitions, but the microwave pulses are disabled (TTL control is turned off). This procedure is used for the normalization of the PL signals. The third image (B) is the background reference; it follows the same time sequence but does not have the  PL light. It is obtained by keeping the AOM turned off. The purpose of this image is to subtract it from the other two, removing the background light before the normalization. To reduce the fluctuation noise, for each sequence corresponding to a value of $\tau$, we do 250 measurements (full images) for each image type.

Finally, to process the data for each value of $\tau$, we add all the 250 images of each kind and do a pixel-wise normalization following $N=(M-B)/(L-B)$, with the letters representing the respective image matrix. Therefore, $N=N(x,y)$ is a matrix representing the normalized image, with $x$ and $y$ being the pixel coordinates. 

\begin{figure}[tb]
\centering
\includegraphics[scale=0.5]{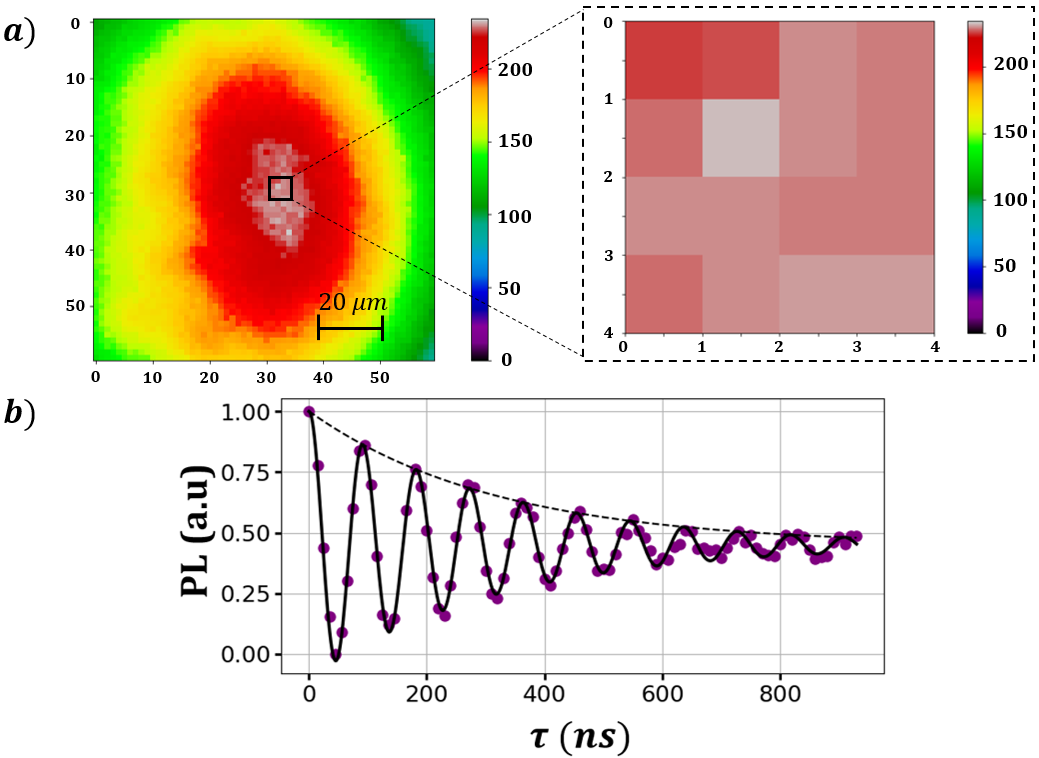}
\caption{a) Fluorescence intensity (a.u) in a M-image. The figure on the right zooms in the 4x4 pixels region indicated on the left. b) Rabi oscillation analysis of the data obtained from the binned 4x4 pixels instead of the whole image. In this case, the coherence time is $\textrm{T}_2^* = 318 \,\ \textrm{ns}$.}
\label{grafrabcrop}
\end{figure}

\section{Results and Discussions}
The Rabi oscillation experiment was performed with a MW power of $40 \,\ \textrm{dBm}$, at $3111.5$ MHz. The PL signal was measured for pulses $\tau$ varying from 0 to 930 ns, using the protocol in the previous section. Notice that each point on the graph seen in Fig. \ref{graftrabi1}(c) corresponds to 37500 microwave pulse sequences (just in the M-image).  

The relaxation constant $\textrm{T}_2^*$ is measured by fitting the data with the Eq. \eqref{rabios}, as in Fig. \ref{graftrabi1}(c). However, it is important to note that in this case the data integrates the PL signal over the whole fluorescing area (laser beam's size $\approx185 \,\mu\textrm{m}$). Therefore, in this case, the measured coherence time is given by an average over a relatively large (field-inhomogeneous) area, as registered in the imaging camera. 
In the literature, this type of measurement typically uses a fast single photodetector, as a photomultiplier or an avalanche photodiode, even when one has a spatially extended ensemble of spins, which effectively integrates the information into a single data point. 

Here, we use a CCD camera instead. To compensate for the much slower acquisition time of this device, we introduced \cite{thesecharlie} the optical imaging protocol described in section \ref{Pulse}. One of the advantages of using a camera is that we have an image (2D map) with relevant information encoded on the spatial distribution of the PL signal. If properly processed, this extra information can be useful.

For example, in Fig. \ref{grafrabcrop}, we use a 4x4 binning of the pixels (1 pixel $\approx 1.67\,\mu\textrm{m}$) from the same image used in Fig. \ref{graftrabi1} ($60\times60$ pixels), and we see that the Rabi oscillations analysis in Fig. \ref{grafrabcrop}(b) results in a significantly greater coherence time, equal to $\text{T}_2^*= 318 \,\ \textrm{ns}$.
This large difference in coherence time may result from several factors, as mentioned in section \ref{spincohe}, including the inhomogeneity of the field produced by the microwave antenna itself, which is the main factor here. 

In this preliminary study, we use the own characteristics of our setup to illustrate how the methods described here can be used to build 2D maps of spin coherence in an ensemble of NV centers. For that, we exploit the physical size of our system that, due to the relative size of the MW antenna and its distance to the sampled (imaged) region, results in an unequal MW excitation of the sample. This is caused by the spatially varying intensity and direction of the B-field, which generates regions of inhomogeneous detuning of the spins in the sample, leading to a large spatial variation of the coherence time $\textrm{T}_2^*$.

\begin{figure}[tb]
\centering
\includegraphics[scale=0.52]{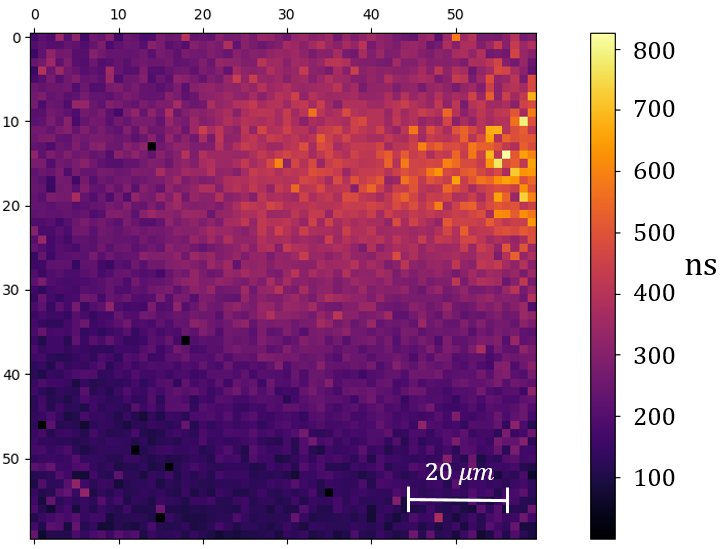}
\caption{Measured spin coherence map of the NV centers. An automated Rabi analysis of each image pixel makes it possible to map the spatial variation of $\textrm{T}_2^*$. The coherence time in the upper right corner exceeds 800 ns. It corresponds to a region near the center of the MW antenna.}
\label{grafmap}
\end{figure}

To illustrate this point, we do a pixel-by-pixel analysis using the data collected with our protocol. The result is shown in Fig. \ref{grafmap}, with a 2D map of the spin coherence of the NV centers in our device. The pixels in the image are color coded with their respective $\textrm{T}_2^*$. One can notice that the coherence times are much longer in the upper right corner, exceeding 800 ns, values much closer to the measured $T_2$ for this sample \cite{thesecharlie,lucas2021,disertmestre}. Not by coincidence, this region is near the center of the antenna's ``loop", where the MW magnetic field is more homogeneous, thus increasing the coherence value.

\bigskip\bigskip
\section{Conclusions}
In this work, we show how to build a 2D map of spatially varying spin coherence times in an engineered device of ultra-pure diamond with NV centers. Our protocol combines optical imaging with customized microwave pulse sequences and automated offline processing to analyze the coherence times pixel by pixel. It uses simply a regular CCD camera to collect the PL signal with spatially encoded spin-coherence information, instead of more sophisticated (and expensive) fast photodetectors, which would still require elaborate scanning methods to build equivalent images.

Here, as a preliminary study, we focused on demonstrating the first proof of concept, exploiting the characteristics of our setup with inhomogeneous B-fields produced by the microwave antenna. However, the ideas and protocol presented here can easily be extended to applications benefiting from this type of sub-micrometric sensor. In fact, this simple and automated pixel-by-pixel analysis is interesting for gathering information about the magnetic field's inhomogeneity through the sample, thus making the NV center a practical sensor of spatial spin-coherence amendable to many future applications.

\section*{Acknowledgment}
We are very thankful to Victor Acosta for providing the sample used in this study and Thalyta Martins for helpful discussions and assistance with the automated data analysis. Authors acknowledge financial support from CAPES (process 88887.372074/2019-00), CNPq (process 141453/2021-4), and FAPESP (grants 2019/27471-0 and 2013/07276-1).


\begin{thebibliography}{18}

\bibitem{acosta2009diamonds}
Victor~M Acosta, Erik Bauch, Micah~P Ledbetter, Charles Santori, K-MC Fu,
  Paul~E Barclay, Raymond~G Beausoleil, H{\'e}lo{\"\i}se Linget, Jean~Francois
  Roch, Francois Treussart, et~al.
\newblock Diamonds with a high density of nitrogen-vacancy centers for
  magnetometry applications.
\newblock {\em Physical Review B}, 80(11):115202, 2009.

\bibitem{nanoscale}
Gopalakrishnan Balasubramanian, I.~Y. Chan, Roman Kolesov, Mohannad Al-Hmoud,
  Julia Tisler, Chang Shin, Changdong Kim, Aleksander Wojcik, Philip~R. Hemmer,
  Anke Krueger, Tobias Hanke, Alfred Leitenstorfer, Rudolf Bratschitsch, Fedor
  Jelezko, and J{\"{o}}rg Wrachtrup.
\newblock Nanoscale imaging magnetometry with diamond spins under ambient
  conditions.
\newblock {\em Nature}, 455(7213):648, 2008.

\bibitem{biomag}
Liam~P McGuinness, Yuling Yan, Alastair Stacey, David~A Simpson, Liam~T Hall,
  Dougal Maclaurin, Steven Prawer, P~Mulvaney, J~Wrachtrup, F~Caruso, et~al.
\newblock Quantum measurement and orientation tracking of fluorescent
  nanodiamonds inside living cells.
\newblock {\em Nature nanotechnology}, 6(6):358--363, 2011.

\bibitem{ultrasensitive}
Liam~T Hall, Charles~D Hill, Jared~H Cole, and Lloyd~CL Hollenberg.
\newblock Ultrasensitive diamond magnetometry using optimal dynamic decoupling.
\newblock {\em Physical Review B}, 82(4):045208, 2010.

\bibitem{Tetienne2017}
Jean-Philippe Tetienne, Nikolai Dontschuk, David~A. Broadway, Alastair Stacey,
  David~A. Simpson, and Lloyd C.~L. Hollenberg.
\newblock Quantum imaging of current flow in graphene.
\newblock {\em Science Advances}, 3(4), 2017.

\bibitem{electric}
Florian Dolde, Helmut Fedder, Marcus~W Doherty, Tobias N{\"o}bauer, Florian
  Rempp, Gopalakrishnan Balasubramanian, Thomas Wolf, Friedemann Reinhard,
  Lloyd~CL Hollenberg, Fedor Jelezko, et~al.
\newblock Electric-field sensing using single diamond spins.
\newblock {\em Nature Physics}, 7(6):459--463, 2011.

\bibitem{nanothermo}
Philipp Neumann, Ingmar Jakobi, Florian Dolde, Christian Burk, Rolf Reuter,
  Gerald Waldherr, Jan Honert, Thomas Wolf, Andreas Brunner, Jeong~Hyun Shim,
  et~al.
\newblock High-precision nanoscale temperature sensing using single defects in
  diamond.
\newblock {\em Nano Letters}, 13(6):2738--2742, 2013.

\bibitem{thermometer2020}
Joonhee Choi, Hengyun Zhou, Renate Landig, Hai-Yin Wu, Xiaofei Yu, Stephen~E
  Von~Stetina, Georg Kucsko, Susan~E Mango, Daniel~J Needleman, Aravinthan~DT
  Samuel, et~al.
\newblock Probing and manipulating embryogenesis via nanoscale thermometry and
  temperature control.
\newblock {\em Proceedings of the National Academy of Sciences},
  117(26):14636--14641, 2020.

\bibitem{quantum}
MV~Gurudev Dutt, L~Childress, L~Jiang, E~Togan, J~Maze, F~Jelezko, AS~Zibrov,
  PR~Hemmer, and MD~Lukin.
\newblock Quantum register based on individual electronic and nuclear spin
  qubits in diamond.
\newblock {\em Science}, 316(5829):1312--1316, 2007.

\bibitem{charlie2021}
Charlie {Oncebay Segura} and Sergio~R. Muniz.
\newblock Diamond-based optical vector magnetometer.
\newblock In {\em 2021 SBFoton International Optics and Photonics Conference
  (SBFoton IOPC)}, São Carlos, Brazil, May 2021.
\newblock \href{https://doi.org/10.1109/sbfotoniopc50774.2021.9461950}{https://doi.org/10.1109/sbfotoniopc50774.2021.9461950}

\bibitem{lucas2021}
Lucas Nunes~Sales de~Andrade, Charlie~Oncebay Segura, and Sérgio~Ricardo
  Muniz.
\newblock Measurements of spin-coherence in nv centers for diamond-based
  quantum sensors.
\newblock In {\em 2021 SBFoton International Optics and Photonics Conference
  (SBFoton IOPC)}, pages 1--4, May 2021.
\\ \href{https://doi.org/10.1109/SBFotonIOPC50774.2021.9461941}{https://doi.org/10.1109/SBFotonIOPC50774.2021.9461941}

\bibitem{Haque-Sumaiya}
Ariful Haque and Sharaf Sumaiaya.
\newblock An overview on the formation and processing of nitrogen-vacancy
  photonic centers in diamond by ion implantation.
\newblock {\em Journal of Manufacturing and Material Processing}, 1(1):6, 2017.

\bibitem{thesepreez}
L.~Du Preez.
\newblock {\em Electron paramagnetic resonance and optical investigations of
  defect centre in diamond}.
\newblock Ph.D. thesis, University of Witwaters, 1965.

\bibitem{doherty2013nitrogen}
Marcus~W Doherty, Neil~B Manson, Paul Delaney, Fedor Jelezko, J{\"o}rg
  Wrachtrup, and Lloyd~CL Hollenberg.
\newblock The nitrogen-vacancy colour centre in diamond.
\newblock {\em Physics Reports}, 528(1):1--45, 2013.

\bibitem{Acosta_PRB}
V.~M. Acosta, A.~Jarmola, E.~Bauch, and D.~Budker.
\newblock Optical properties of the nitrogen-vacancy singlet levels in diamond.
\newblock {\em Phys. Rev. B}, 82:201202, 2010.

\bibitem{thesecharlie}
Charlie Oncebay Segura.
\newblock {\em Diamond studies for applications in quantum technologies}.
\newblock PhD thesis, Universidade de São Paulo, Instituto de Física de São
  Carlos, 2019.
\\ \href{https://doi.org/10.11606/T.76.2019.tde-01082019-152208}{https://doi.org/10.11606/T.76.2019.tde-01082019-152208}

\bibitem{ultralong}
Erik Bauch, Connor~A Hart, Jennifer~M Schloss, Matthew~J Turner, John~F Barry,
  Pauli Kehayias, Swati Singh, and Ronald~L Walsworth.
\newblock Ultralong dephasing times in solid-state spin ensembles via quantum
  control.
\newblock {\em Physical Review X}, 8(3):031025, 2018.

\bibitem{disertmestre}
Lucas Nunes~Sales de~Andrade.
\newblock Optical measurements of electron spin coherence of nitrogen-vacancy
  centers in diamond.
\newblock M.S. thesis, Instituto de Física de São Carlos, Universidade
  de S{\~a}o Paulo, São Carlos, 2021.
\newblock \href{https://doi.org/10.11606/D.76.2021.tde-20092021-121459}{https://doi.org/10.11606/D.76.2021.tde-20092021-121459}

\end{thebibliography}
\end{document}